\documentclass[9pt,conference]{IEEEtran}

\usepackage[preprint]{waspaa25}

\usepackage{bm} 
\usepackage{hyperref}
\usepackage{booktabs}
\usepackage{multirow}
\usepackage{siunitx}
\usepackage{tikz}

\usetikzlibrary{positioning}

\newcommand{\RR}{\mathbb{R}}
\newcommand{\x}{\mathbf{x}}
\newcommand{\y}{\mathbf{y}}
\newcommand{\barx}{\bar{\x}}
\newcommand{\bary}{\bar{\y}}
\newcommand{\barz}{\bar{\mathbf{z}}}
\newcommand{\tildey}{\tilde{\y}}
\newcommand{\tildex}{\tilde{\x}}
\newcommand{\tildez}{\tilde{\mathbf{z}}}
\newcommand{\btheta}{\bm{\theta}}

\title{Improving Inference-Time Optimisation for Vocal Effects Style Transfer with a Gaussian Prior}

\name{Chin-Yun Yu$^{1}$\thanks{This research is supported jointly by UKRI (grant number EP/S022694/1) and QMUL. This research utilised Queen Mary's Apocrita HPC facility, supported by QMUL Research-IT. \href{http://doi.org/10.5281/zenodo.438045}{doi: 10.5281/zenodo.438045}},
      Marco A. Martínez-Ramírez$^{2}$,
      Junghyun Koo$^{2}$,
      Wei-Hsiang Liao$^{2}$,
      Yuki Mitsufuji$^{2,3}$,
      Gy\"orgy Fazekas$^{1}$}
\address{$^{1}$Centre for Digital Music, Queen Mary University of London, London, UK \\
$^{2}$Sony AI, Tokyo, Japan\; $^{3}$Sony Group Corporation, Tokyo, Japan
}

\begin{document}

\maketitle

\begin{abstract}
    Style Transfer with Inference-Time Optimisation (ST-ITO) is a recent approach for transferring the applied effects of a reference audio to an audio track.
    It optimises the effect parameters to minimise the distance between the style embeddings of the processed audio and the reference.
    However, this method treats all possible configurations equally and relies solely on the embedding space, which can result in unrealistic configurations or biased outcomes.
    We address this pitfall by introducing a Gaussian prior derived from the DiffVox vocal preset dataset over the parameter space.
    The resulting optimisation is equivalent to maximum-a-posteriori estimation.
    Evaluations on vocal effects transfer on the MedleyDB dataset show significant improvements across metrics compared to baselines, including a blind audio effects estimator, nearest-neighbour approaches, and uncalibrated ST-ITO.
    The proposed calibration reduces the parameter mean squared error by up to 33\% and more closely matches the reference style.
    Subjective evaluations with 16 participants confirm the superiority of our method in limited data regimes.
    This work demonstrates how incorporating prior knowledge at inference time enhances audio effects transfer, paving the way for more effective and realistic audio processing systems.
\end{abstract}

\section{Introduction}
\label{sec:intro}

Audio production is a complex task that needs years of practice and experience to master.
Audio engineers need to know their tools and understand their working context, such as the instrument's characteristics, the recording environment, and the artist's intention.
For the latter, it is common to use a reference track provided by the client to guide the process \cite{vanka2024the}.
This user-friendly paradigm has also been applied to automatic music mixing with deep learning \cite{martinez2022automatic, steinmetz2022style}.
The systems can either be a transformation model that directly manipulates the audio \cite{martinez2022automatic, koo_music_2023} or predict the control parameters of audio effects \cite{steinmetz_automatic_2020,steinmetz2022style,vanka_diff-mst_2024}, where the latter is more preferable since it is more controllable and generalisable to real-world workflow.
At the same time, the high degree of freedom of the former could introduce artefacts.
Nevertheless, these parameter-based systems use fixed effects and track routings, which hinder their flexibility.

\subsection{Inference-time optimisation (ITO)}
\label{ssec:ito}

More recent works explore the idea of processing the audio during inference time, which is free from training conditions and can generalise to arbitrary effects and routings.
Text2Fx \cite{chu2025text2fx} leverages the joint text and audio embedding space of CLAP \cite{elizalde2023clap} to optimise the parameters of an equaliser and a reverb so the processed audio has an embedding close to the text prompt.
Steinmetz et al. \cite{christian_j_steinmetz_2024_14877423} proposed Style Transfer with Inference Time Optimisation (ST-ITO), which includes a self-supervised method to construct a mixing style encoder and a non-differentiable way to find suitable parameters that have the minimum embedding distance to the reference audio.
With their proposed encoder AFx-Rep, ST-ITO achieves state-of-the-art performance on this task.
Koo et al. \cite{koo2025ito-m} proposed ITO-Master, which uses training-free audio features from \cite{vanka_diff-mst_2024} as the style encoder.
Our work extends ITO by addressing the pitfall of solely minimising the embedding distance, which neglects the prior information of the effect parameters.

\begin{figure}[t]
    \centering
    \includegraphics[clip, trim=5mm 5mm 1.5cm 10mm, width=0.9\columnwidth]{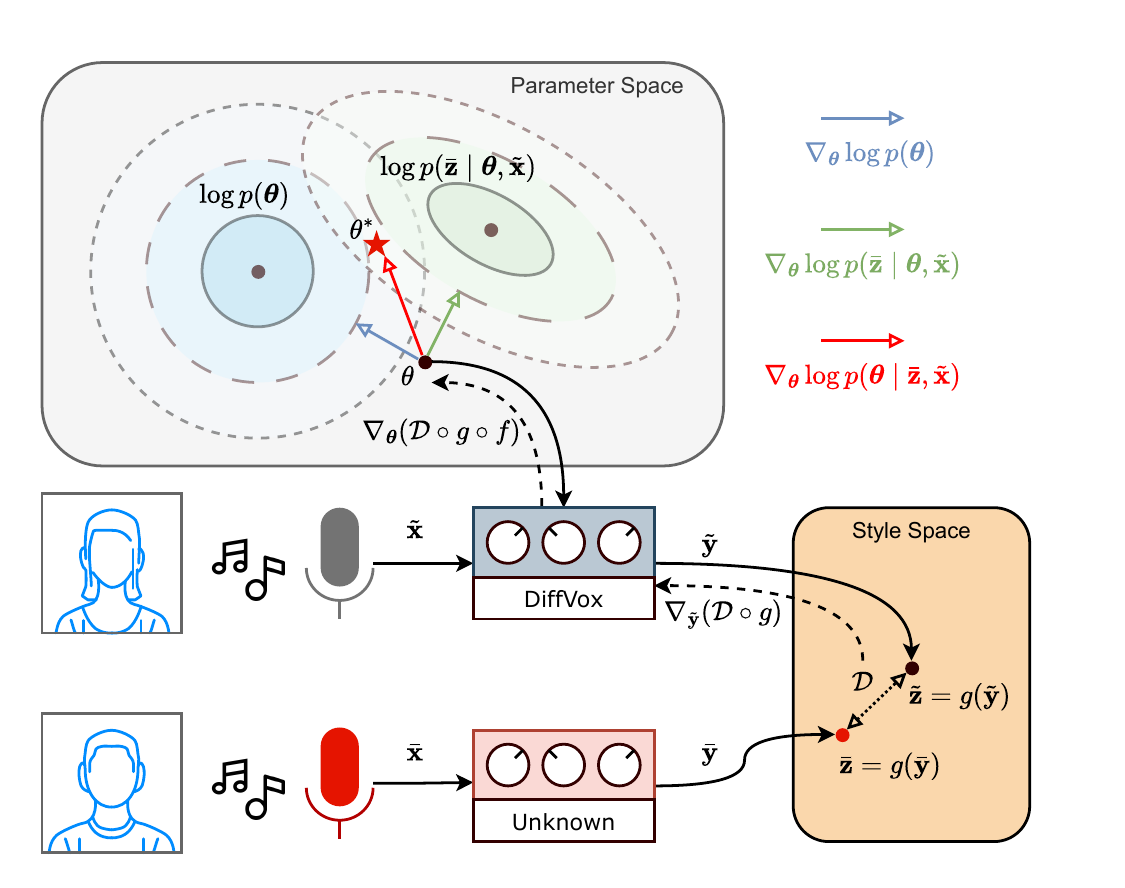}
    \caption{Overview of the proposed calibration method.
        The prior and the likelihood log-probability densities in the parameter space are represented by two concentric ellipses, coloured blue and green, respectively. Darker colours indicate higher density. The coloured arrows indicate the gradients of the log-probability densities. The red star is the optimal parameters $\btheta^*$ for the vocal effects style transfer.}
    \label{fig:overview}
\end{figure}

\subsection{Problem statement}
\label{sec:problem}
Given a reference track $\bary \in \RR^N$, $C$ raw tracks $\tildex \in \RR^{N \times C}$, and a content-invariant style encoder $g: \RR^N \to S^{D-1}$ which maps the audio into a $D$-dimensional unit vector, we would like to process the tracks in a way that the resulting $\tildey$ has the same effects applied as $\bary$.
Here, \emph{raw track} means that no or very few effects have been applied.
A natural criterion is to make the two style embeddings $\tildez = g(\tildey)$ and $\barz = g(\bary)$ as close as possible \cite{chu2025text2fx,christian_j_steinmetz_2024_14877423}.
We further assume that the reference is the result of applying an effects model $f: \RR^{N \times C} \times \RR^M \to \RR^N$ with parameters $\btheta \in \RR^M$, to its corresponding raw tracks $\barx \in \RR^{N \times C}$, which we do not have access to.
Assuming the effects configuration encompasses what \emph{mixing style} is.
Transferring the reference style becomes a problem of estimating the parameters $\btheta$ given $\barz$ and $\tildex$, i.e., the posterior distribution $p(\btheta \mid \barz, \tildex)$.
Once the parameters are estimated, we can apply the effects model to $\tildex$ to obtain the desired output $\tildey = f(\tildex, \btheta)$.

In ST-ITO, the style distance is the cosine distance $\mathcal{D}(\tildez, \barz) = 1 - \tildez^\top \barz$.
The authors find the optimal parameters by $\btheta^* = \arg\min_{\btheta} \mathcal{D}(\tildez, \barz)$.
We could view this as maximising the likelihood $p(\barz \mid \btheta, \tildex)$.
Nevertheless, according to Bayes' rule, this is only proportional to the posterior $p(\btheta \mid \barz, \tildex) \propto p(\barz \mid \btheta, \tildex) p(\btheta \mid \tildex)$ when the prior $p(\btheta \mid \tildex)$ is uniform over the parameter space.
Or, when the likelihood function is a Dirac delta function, the prior does not affect the posterior.
This could happen if $g$ is a perfect encoder, i.e., injective to style (a.k.a $\btheta$) and fully represents the entire parameter space, which is, unfortunately, not the case in practice.
Thus, a reasonable prior is needed to guide the search for the parameters $\btheta$.
Such prior tells us that some configurations are more likely to happen than others, thus reducing the risk of arriving at an unrealistic solution.
This prior can come from heuristics, such as the knowledge of the audio engineer, or from data-driven methods, which is the focus of this work.

\subsection{Related works}
\label{ssec:related}

The concept of utilising external guidance during inference time is not new and has been extensively studied in the field of generative modelling \cite{manilow2022source}, especially using diffusion models \cite{sohl-dickstein_deep_2015, song2021scorebased,ho2020denoising}.
Diffusion models learn the gradients of the log-probability density $\nabla_{\btheta} \log p(\btheta)$ that can be used to draw samples from the distribution $p(\btheta)$.
(We omit the condition variable $t$ representing diffusion time steps for demonstration purposes.)
Here, $\btheta$ could be any targeted variable.
Using the equivalence of $\nabla_{\btheta} \log p(\btheta \mid \barz) = \nabla_{\btheta} \log p(\barz \mid \btheta) + \nabla_{\btheta} \log p(\btheta)$, one can design a likelihood function $p(\barz \mid \btheta)$ on a given condition $\barz$ thereby turning the unconditional generative model into a conditional one \cite{song2021scorebased, chung2023diffusion}.
This is known as \emph{posterior sampling}.
This idea has been successfully applied in solving inverse problems such as image and audio restoration \cite{kawar2022denoising, yu2023conditioning, hernandez-olivan_vrdmg_2023, Lemercier2024}, and source separation \cite{mariani2024multi, hirano_diffusion-based_2023}.
Recently, a similar approach has been used to control music generation models known as Diffusion Inference-Time T-Optimisation (DITTO) \cite{novack2024ditto,zachary_novack_2024_14877469}.
The difference is that DITTO treats the whole sampling process as a function and backpropagates the gradients through it to optimise the initial latent.
This, however, deviates from posterior sampling and is more similar to the ITO methods we mentioned in \cref{ssec:ito}, in which the sampling process is similar to $f$ and the initial latent is similar to $\btheta$.
(This can be corrected by regularising the initial latent to its initial distribution $p(\btheta) = \mathcal{N}(0, \mathbf{I})$ during the optimisation, similar to our approach proposed in \cref{ssec:map}.)

Posterior sampling has a strong prior that guarantees correctness, which the previous ITO-based style transfer works do not have.
The reason is apparent: large-scale effect parameter datasets are needed to capture their distribution, which are rare and expensive to collect.
Our work does not address this issue, but instead focuses on using a limited amount of data to showcase that missing prior information can lead to suboptimal results and draw more attention to the importance of audio effects prior.
Our contributions are as follows:
1) converting the effects style transfer into a maximum-a-posteriori estimation problem,
2) improvements in vocal effects style transfer, outperforming baselines objectively and subjectively using a Gaussian prior and a differentiable effects model (DiffVox),
3) exploration of multiple encoders for style representation, and
4) scalability to limited paired data regimes.
We also open-source our implementation and experiments on GitHub~\footnote{\href{https://github.com/SonyResearch/diffvox}{github.com/SonyResearch/diffvox}}.

\section{Methodology}
\label{sec:method}

\subsection{The maximum-a-posteriori estimation}
\label{ssec:map}
To correct the biased posterior of ST-ITO, we propose introducing a non-uniform prior $p(\btheta \mid \tildex)$ on the parameters $\btheta$.
We further assume that the prior is independent of the raw vocal $\tildex$, so we can write it as $p(\btheta \mid \tildex) = p(\btheta)$ to simplify the problem.
The optimal parameters $\btheta^*$ thus equals the following \emph{maximum-a-posteriori} (MAP) estimation:
\begin{equation}
    \label{eq:map}
    \btheta^* = \underset{\btheta}{\arg\max} \log p(\barz \mid \btheta, \tildex) + \alpha \log p(\btheta)
\end{equation}
where $\alpha$ is a hyperparameter that controls the prior strength.
\Cref{fig:overview} gives an intuitive illustration of \cref{eq:map} in a 2-dimensional parameter space.
The green arrow is the gradient of the log-likelihood, which is used in ST-ITO \cite{christian_j_steinmetz_2024_14877423}.
The red arrow is the optimisation direction we proposed, which is the sum of the gradients of the two distributions as mentioned in \cref{ssec:related}.
To define the likelihood $p(\barz \mid \btheta, \tildex)$, we first convert the distance measure $\mathcal{D}:\RR^D \times \RR^D \to [0, 2]$ into a probability density function as:
\begin{equation}
    p(\barz \mid \tildez) = \frac{1}{\sqrt{2\pi}\sigma}\exp\left( -\frac{\arccos\left(1 - \mathcal{D}(\tildez, \barz)\right)^2}{2\sigma^2}\right)
\end{equation}
where $\sigma^2$ is a hyperparameter that controls the variance of the distribution.
We measure the embedding distance using $\arccos$ since embeddings lie on a (D-1)-sphere and the distance on the surface equals the angle between the two vectors.
We parametrise the likelihood to be $p(\barz \mid \btheta, \tildex) = p(\barz \mid \tildez = (g \circ f)(\tildex, \btheta))$.

\subsection{The effects model and Gaussian prior}
\label{ssec:effects}

Due to the lack of multi-track effects datasets, we demonstrate our approach on the simpler case of processing a single track.
In this case, we utilise the DiffVox dataset \cite{diffvox}, which comprises 365 vocal presets derived from professionally mixed songs, thereby providing high-quality effect presets to build the prior.
The differentiable effects model $f: \RR^{N \times 1} \times \RR^M \to \RR^{N \times 2}$ converts a mono vocal track into a stereo track and has $M=130$.
$f$ consists of a six-band parametric equaliser (PEQ), a dynamic range controller (DRC), a ping-pong delay, a feedback delay network (FDN) reverb, and a panner.
The routing of the effects is shown in \cref{fig:effects-chain}.

\begin{figure}[h]
    \centering
    \resizebox{!}{!}{
        \begin{tikzpicture}[
                auto,
                block/.style={rectangle, draw, thick, fill=white, minimum width=1.0cm, minimum height=0.5cm, text centered},
                sum/.style={circle, draw, thick, inner sep=0pt, minimum size=0.2cm},
                line/.style={draw, -latex},
                node distance=0.2cm and 1.0cm,
            ]

            \node [sum] (input) {};
            \node [block] (peq) [right=of input] {PEQ};
            \node [block] (drc) [right=of peq] {DRC};
            \node [block] (pan) [right=of drc] {Panner};
            \node [block] (dly) [below=of pan] {Delay};
            \node [block] (rev) [below=of dly] {Reverb};
            \node [sum, name=output, right=of pan] {};

            \draw[thick, ->] (input.east) -- (peq.west);
            \draw[thick, ->] (peq.east) -- (drc.west);
            \draw[thick, ->] (drc.east) -- (pan.west);
            \draw[thick, ->] (drc.east) -- (dly.west);
            \draw[thick, ->] (drc.east) -- (rev.west);
            \draw[thick, ->] (dly.south) -- (rev.north);
            \draw[thick, ->] (pan.east) -- (output.west);
            \draw[thick, ->] (dly.east) -- (output.west);
            \draw[thick, ->] (rev.east) -- (output.west);

        \end{tikzpicture}}
    \caption{The effects chain in DiffVox.}
    \label{fig:effects-chain}
\end{figure}

The vocal presets $\bm{\Theta} \in \RR^{130 \times 365}$ were retrieved by fitting $f$ to paired dry and wet vocal tracks from a private multi-track dataset, which contains mostly mainstream Western music, with \SI{44.1}{\kHz} sampling rate.
Each track is split into overlapping long chunks and silent chunks are dropped.
The effects are trained on each track for 2,000 steps using Adam \cite{kingma_adam} with a learning rate of 0.01 and a maximum batch size of 35.
The training loss is a combination of multi-scale STFT (MSS) losses \cite{wang2018neural} and mean absolute error (MAE) loss on microdynamics \cite{nercessian_direct_2022} in two different scales (MLDR) on the left, right, mid, and side channels.
The MSS is the sum of spectral convergence and the MAE of log-magnitude spectrograms computed using \texttt{auraloss}\footnote{\href{https://github.com/csteinmetz1/auraloss}{github.com/csteinmetz1/auraloss}}.
The spectrograms for MSS are computed with FFT sizes of $\{128, 512, 2048\}$ and \SI{75}{\percent} overlap.

We initially attempted to fit a diffusion model to the presets; however, due to data scarcity, the model failed to generate meaningful presets.
Thus, we propose a simple Gaussian prior $p(\btheta) = \mathcal{N}(\bar{\btheta}, \bm{\Sigma}_{\btheta})$ instead,
using the sample mean $\bar{\btheta} = \frac{1}{365} \sum_{i=1}^{365} \bm{\Theta}_i$ and the sample covariance $\bm{\Sigma}_{\btheta} = \frac{1}{364} \sum_{i=1}^{365} (\bm{\Theta}_i - \bar{\btheta})(\bm{\Theta}_i - \bar{\btheta})^\top$ as parameters of the prior.
Although the statistical tests in \cite{diffvox} indicate that the presets are not multivariate normal, we argue that this weak model is still beneficial, as it captures the non-uniformity to some extent.

\subsection{Working with multiple vocals and channels}
\label{ssec:multiple}

The effects model of DiffVox $f$ produces stereo vocals due to the spatial effects.
Let us denote its two-channel output as $\begin{bmatrix}\tildey_\text{l} \; \tildey_\text{r}\end{bmatrix} = f(\tildex, \btheta)$.
Following ST-ITO \cite{christian_j_steinmetz_2024_14877423}, we compute four style embeddings $\barz_\text{m}$, $\barz_\text{s}$, $\tildez_\text{m}$, and $\tildez_\text{s}$ from the mid and side channels of the vocals as $\begin{bmatrix}\tildey_\text{m} \; \tildey_\text{s}\end{bmatrix} = \begin{bmatrix}\tildey_\text{l} + \tildey_\text{r} \; \tildey_\text{l} - \tildey_\text{r}\end{bmatrix}$ and $\begin{bmatrix}\bary_\text{m} \; \bary_\text{s}\end{bmatrix} = \begin{bmatrix}\bary_\text{l} + \bary_\text{r} \; \bary_\text{l} - \bary_\text{r}\end{bmatrix}$.
We assume each channel's likelihood is independent of each other, thus
$p(\barz_\text{m}, \barz_\text{s} \mid \btheta, \tildex) = p(\barz_\text{m} \mid \tildez = \tildez_\text{m}) p(\barz_\text{s} \mid \tildez = \tildez_\text{s})$.
In addition, we may have multiple reference embeddings $\bar{\mathbf{Z}} = \{(\barz_{\text{m}_1}, \barz_{\text{s}_1}), (\barz_{\text{m}_2}, \barz_{\text{s}_2}), \ldots\}$ and multiple raw vocals $\tilde{\mathbf{X}} = \{\tildex_1, \tildex_2, \ldots\}$.
In this case, we replace $\log p(\barz \mid \btheta, \tildex)$ in \cref{eq:map} with its expectation, which is
\begin{multline}
    \label{eq:exp_stereo_multi}
    \mathbb{E}_{\bar{\mathbf{Z}}, \tilde{\mathbf{X}}} \left[\log p(\barz_\text{m}, \barz_\text{s} \mid \btheta, \tildex) \right] \\
    = \frac{1}{|\tilde{\mathbf{X}}||\bar{\mathbf{Z}}|} \sum_{(\barz_\text{m}, \barz_\text{s}) \in \bar{\mathbf{Z}}} \sum_{\tildex \in \tilde{\mathbf{X}}} \log p(\barz_\text{m}, \barz_\text{s} \mid \btheta, \tildex).
\end{multline}
We set $\sigma_\text{m}^2$ to be the average of the squared term in the exponent of the likelihood on the mid channel:
\begin{equation}
    \sigma_\text{m}^2 = \frac{1}{|\bar{\mathbf{Z}}_\text{m}||\tilde{\mathbf{Z}}_\text{m}|} \sum_{\barz \in \bar{\mathbf{Z}}_\text{m}} \sum_{\tildez \in \tilde{\mathbf{Z}}_\text{m}} \arccos\left(1 - \mathcal{D}\left(\tildez, \barz\right)\right)^2
\end{equation}
so \cref{eq:exp_stereo_multi} is always maximised with respected to $\sigma_\text{m}^2$ \cite{asperti_balancing_2020}.
$\tilde{\mathbf{Z}}_\text{m}$ and $\bar{\mathbf{Z}}_\text{m}$ are the style embeddings for the mid channel.
The same applies to the side channel with embeddings $\tilde{\mathbf{Z}}_\text{s}$ and $\bar{\mathbf{Z}}_\text{s}$ to calculate $\sigma_\text{s}^2$.

\section{Experimental Details}
\label{sec:exp}
We evaluate the proposed method on 70 vocal tracks from the MedleyDB dataset \cite{bittner2014medleydb,bittner2016medleydb}, with presets derived in the same manner as described in \cref{ssec:effects}, denoted as \textbf{oracle}.
Oracle defines the upper bound of $f$ modelling ability.
We test three different encoders $g$: the AFx-Rep encoder from ST-ITO \cite{christian_j_steinmetz_2024_14877423} and two signal processing-based encoders.
The AFx-Rep encoder is based on the PANNs architecture \cite{kong2020panns} and was trained to identify the applied effects and their configurations from a randomly processed audio.
The second encoder is 25 Mel-frequency cepstral coefficients (MFCCs) computed from 128 Mel bands.
The last encoder is based on standard MIR features, which consist of root-mean-square (RMS) energy, crest factor, dynamic spread, spectral centroid, spectral flatness, and spectral bandwidth \cite{brecht_de_man_2014_1416832, ma2015intelligent}.
The MFCCs and MIR features are computed using a 2048 frame length with \SI{50}{\percent} overlap, and the Hanning window is used for the spectrograms.
Finally, the features are reduced to one vector by computing the mean, standard deviation, skewness, and kurtosis along the time axis.
We apply normalisation $\x \mapsto \x/\|\x\|_2$ to all the embeddings.

To objectively evaluate performance, we design the following experiment similar to the self-supervised training process in \cite{steinmetz2022style, vanka_diff-mst_2024,koo2025ito-m}.
We split the vocal tracks into non-overlapping eleven-second segments and select active segments in which at least \SI{50}{\percent} of the processed audio is over \SI{-60}{\decibel}.
We then randomly split the segments equally into two sets, denoted as A and B.
The embeddings of A's processed audio and B's raw vocal are used as $\barz$ and $\tildex$, respectively.
After estimating the parameter $\btheta^*$, we directly compare the resulting $\tildey$ with the processed audio of B.
We drop five tracks due to the insufficient number of segments.
We minimise the negative log-posterior \cref{eq:map} using Adam \cite{kingma_adam} for 1,000 steps with a learning rate of 0.01 to reach sufficient convergence, similar to the variational posterior sampling \cite{mardani2024a}.
These parameters were decided based on early trial runs.
The parameters are initialised to the mean of the presets $\bar{\btheta}$.

We compare the proposed method with the following baselines:
\begin{itemize}
    \item \textbf{Mean}: The presets' mean $\bar{\btheta}$, which is the mode of the prior $\bar{\btheta} = \arg\max_{\btheta} p(\btheta)$ and the solution of \cref{eq:map} when $\alpha = \infty$.
    \item \textbf{Regression}: A neural network that performs blind estimation of the parameters (equals the posterior $p(\btheta \mid \bary)$), trained with a mean squared error (MSE) loss on the parameters.
    \item \textbf{Nearest neighbour (NN)}: It finds the nearest preset from the training set as the parameters. We examine four different spaces to measure the distance: the parameter space ($\btheta$) and the embedding space of the three encoders. We pick the preset nearest to the oracle parameters when using the parameter space.
\end{itemize}
To train the regression baseline, we split vocals into non-overlapping ten-second segments, and we select active segments in which at least \SI{80}{\percent} of the audio is over \SI{-60}{\decibel}.
We select 17 out of 365 tracks for validation and the remaining tracks for training.
The input features include RMS, crest factor, dynamic spread, and a Mel-spectrogram with 80 Mel bands.
We compute the features with a hop size of 256 on the left, right, mid, and side channels and concatenate them along the channel dimension.
The model has five convolutional layers with channel sizes [512, 512, 768, 1024, 1024] and a kernel size of 5.
Average pooling with a stride of 3, batch normalisation, and ReLU activation are sequentially applied between the layers.
Finally, we use global max-pooling followed by a fully connected layer that predicts the parameters.
The model is trained using AdamW \cite{loshchilov2018decoupled} with a learning rate of 0.001 and a batch size of 64.
Due to the small amount of data (roughly 4177 segments), the model overfits soon after 6,500 steps.
We initially tried processing random raw vocals with random parameters to augment the training data, similar to \cite{peladeau2024blind}.
However, the performance on the validation set is worse, likely due to the intrinsic difference between the DiffVox effects and the actual processing of the data.
Thus, we only use the original processed vocals for training.
When multiple references are given during inference, we average the predictions.

\section{Results}

\begin{table}[h]
    \caption{Median scores of the proposed methods and baselines. The best scores in their respective categories are highlighted in bold.}
    \begin{center}
        \begin{tabular}{llrrrrr}
            \toprule
            \multicolumn{2}{c}{\textbf{Method}} & \multicolumn{2}{c}{\textbf{MSS} $\downarrow$} & \multicolumn{2}{c}{\textbf{MLDR} $\downarrow$} & \textbf{PMSE} $\downarrow$                                        \\
                                                &                                               & l/r                                            & m/s                        & l/r        & m/s        &            \\
            \midrule
            \multicolumn{2}{l}{Oracle}          & 0.775                                         & 1.012                                          & 0.313                      & 0.383      & 0.0                     \\
            \midrule
            \multicolumn{2}{l}{Mean}            & +0.354                                        & +0.836                                         & +0.503                     & +0.692     & +5.310                  \\
            \multicolumn{2}{l}{Regression}      & +0.281                                        & +0.574                                         & +0.480                     & +0.695     & +5.002                  \\
            \midrule
            \multicolumn{2}{l}{NN-$\btheta$}    & +0.381                                        & +0.675                                         & +0.518                     & +0.629     & +\bf 4.145              \\
            \multicolumn{2}{l}{NN-AFx-Rep}      & +0.321                                        & +0.672                                         & +\bf 0.320                 & +\bf 0.504 & +9.463                  \\
            \multicolumn{2}{l}{NN-MFCC}         & +\textbf{0.274}                               & +\bf 0.464                                     & +0.424                     & +0.559     & +8.374                  \\
            \multicolumn{2}{l}{NN-MIR}          & +0.424                                        & +0.803                                         & +0.561                     & +0.706     & +10.019                 \\
            \midrule
            Encoder                             & $\alpha$                                                                                                                                                           \\
            \midrule
            \multirow{4}{*}{AFx-Rep}            & 0.0                                           & +0.435                                         & +0.570                     & +0.343     & +0.424     & +7.756     \\
                                                & 0.01                                          & +0.221                                         & +0.606                     & +\bf 0.249 & +\bf 0.402 & +5.924     \\
                                                & 0.1                                           & +\bf 0.211                                     & +\bf 0.513                 & +0.321     & +0.445     & +\bf 5.168 \\
                                                & 1.0                                           & +0.318                                         & +0.795                     & +0.427     & +0.629     & +5.339     \\
            \midrule
            \multirow{4}{*}{MFCC}               & 0.0                                           & +0.761                                         & +0.897                     & +1.047     & +0.977     & +9.255     \\
                                                & 0.01                                          & +0.507                                         & +0.531                     & +0.720     & +0.765     & +6.706     \\
                                                & 0.1                                           & +0.333                                         & +\bf 0.469                 & +0.514     & +0.621     & +5.661     \\
                                                & 1.0                                           & +\bf 0.312                                     & +0.563                     & +\bf 0.459 & +\bf 0.547 & +\bf 5.250 \\
            \midrule
            \multirow{4}{*}{MIR}                & 0.0                                           & +0.782                                         & +1.105                     & +0.873     & +0.797     & +7.103     \\
                                                & 0.01                                          & +0.598                                         & +1.505                     & +0.856     & +0.854     & +5.622     \\
                                                & 0.1                                           & +0.490                                         & +0.807                     & +0.778     & +0.778     & +5.359     \\
                                                & 1.0                                           & +\bf 0.363                                     & +\bf 0.714                 & +\bf 0.508 & +\bf 0.695 & +\bf 5.319 \\

            \bottomrule
        \end{tabular}
        \label{tab:results}
    \end{center}
\end{table}

\subsection{Objective evaluation}
\label{ssec:objective}
The evaluation metrics are the MSS and MLDR losses given in \cref{ssec:effects} and the MSE loss on the parameters (PMSE).
We use oracle as the target to calculate PMSE since there is no better ground-truth.
The median scores across 65 tracks are shown in \cref{tab:results}.
We report the median due to extreme outliers when evaluating on a few tracks' mid and side channels.
Firstly, examining the nearest neighbour methods (denoted as NN- followed by the embedding name), we see that NN-$\btheta$ performs the best in terms of PMSE, but not for the other metrics.
This implies that having close parameters does not mean their audio features are closer.
NN-MFCC is better suited for MSS, which is expected since the MFCCs primarily encode spectral information, while NN-AFx-Rep is better for MLDR, e.g., dynamics.
Both encoders' PMSEs are way larger than NN-$\btheta$, which could be the misalignment of the parameter and audio feature spaces or the quasi/approximate symmetry in the effects \cite{hayes_equivariant_2025} (i.e., different configurations but similar sound).

We test the proposed method with $\alpha \in \{0.0, 0.01, 0.1, 1.0\}$.
The proposed prior improves the original ST-ITO ($\alpha=0.0$) in all metrics, especially in PMSE, where the best ones are comparable to the mean baseline.
AFx-Rep with $\alpha=0.1$ performs the best and surpasses its NN counterpart in all metrics.
MFCC with $\alpha=1.0$ performs comparably to its NN counterpart and has a lower PMSE.
Both methods outperform the regression model in MSS and MLDR, demonstrating the benefits of calibration in cases with limited paired data and the effectiveness of the Gaussian prior.

\subsection{Subjective listening test}
\label{ssec:subjective}

Based on the objective evaluation, we select the oracle and the four best-performing methods from each category: regression, NN-MFCC, AFx-Rep ($\alpha=0.1$), and MFCC ($\alpha=1.0$), to carry out a MUSHRA-like listening test \cite{series2014method}.
We select one track and its loudest part, lasting 7 to 11 seconds, from each singer in MedleyDB.
We exclude tracks that lack sufficiently long singing phrases, have minimal effects applied, such as classical singing, or feature a distinctly different singing style from the rest.
We then randomly pair two tracks together as one trial.
In each trial, the processed audio of one track is used as $\bary$ and the raw vocal of the other track is used as $\tildex$.
The reference and raw vocals do not have the same singer/recording environment and the available samples are also limited ($|\tilde{\mathbf{X}}| = |\bar{\mathbf{Z}}| = 1$), thus making it more challenging than the experiment in \cref{ssec:objective}.
If the pitch ranges in a pair are too different, we find another part of the same track with similar pitches.
We include the raw vocal $\tildex$ as a hidden low anchor.
Participants are asked to compare $\tildey$ of the five methods and the anchor with the reference $\bary$ and rate how close the applied effects are to the reference on a scale from 0 to 100.
We instruct participants to identify the low anchor and rate it below 20.

We construct the evaluation website using webMUSHRA \cite{webmushra}.
We divided the 19 trials into two sets and randomly assigned them to participants to reduce fatigue.
The sets are divided so that each track appears only once in each set, and if it is used as a reference in one set, it can only be used as the raw vocal in the other set.
The orders of the trials and the methods are randomised.
We posted the link through research communities and social media and have received 26 responses as of this writing.
We examined the histogram of accuracies of picking the low anchor to exclude outliers.
It shows a bimodal distribution with modes at both ends.
Therefore, we use \SI{40}{\percent} as the threshold, which perfectly filter out the low-performance group with eight responses.
Then, we check the variation in ratings across trials for each participant and exclude two more participants with variations that appear to be larger than the rest.
Finally, we average the ratings across trials for each participant.

\begin{figure}[h]
    \centering
    \includegraphics[width=0.95\columnwidth]{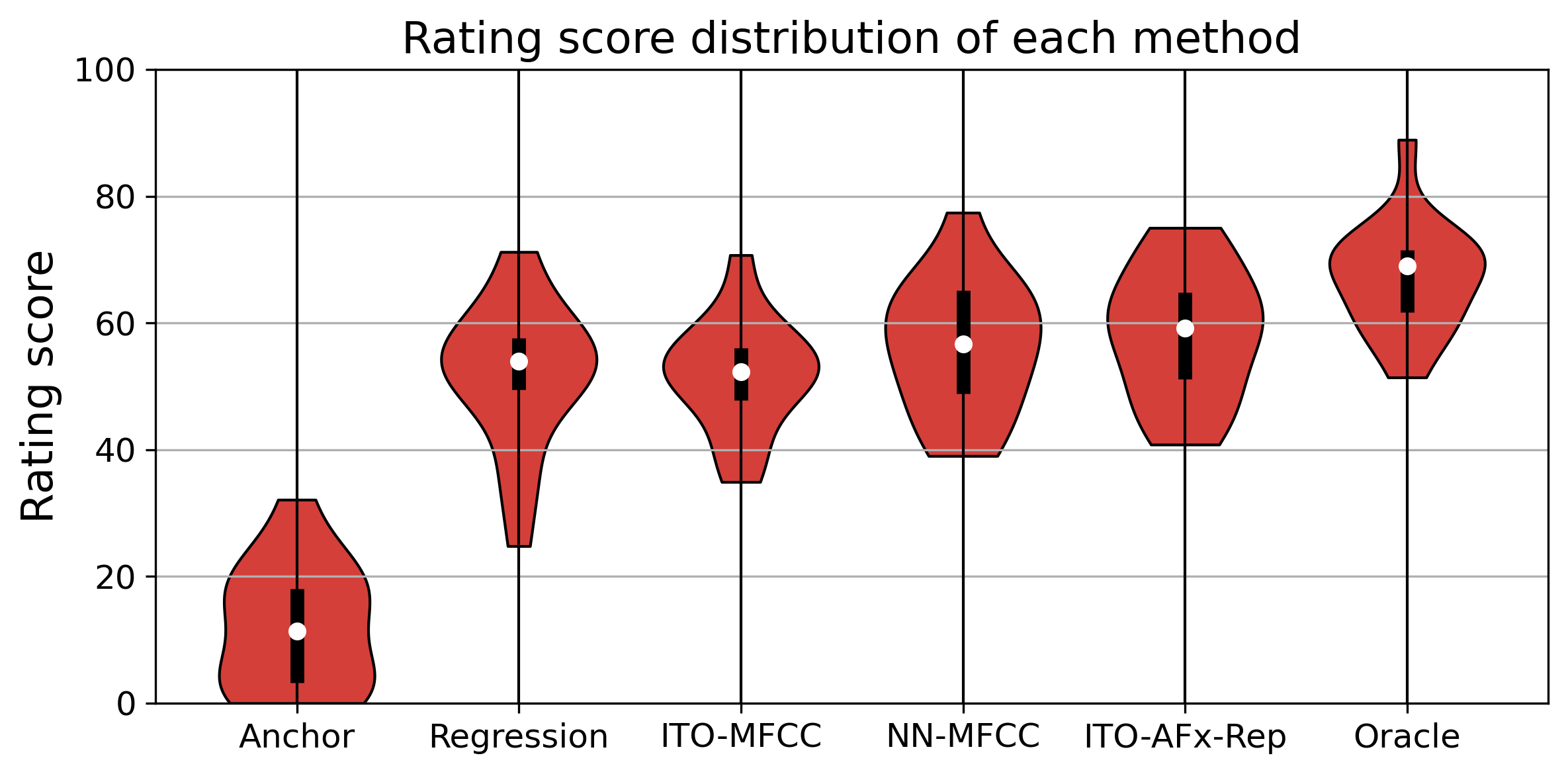}
    \caption{Violin plot of the average ratings sorted based on the mean. The white dot is the median, and the black thick lines are the interquartile range. }
    \label{fig:mushra}
\end{figure}

The distributions of average ratings are shown in \cref{fig:mushra}.
The oracle is not rated close to 100, which is expected due to the difference between the fitted effects, the reference's actual effects, and the singer's timbre.
The regression model is rated the lowest and AFx-Rep is rated the highest on average, though the difference between the two is not statistically significant using the Wilcoxon signed-rank test \cite{wilcoxon1945individual} ($p=.117$).
In summary, the subjective evaluation indicates that the proposed method, combined with the AFx-Rep encoder, tends to transfer the reference style more effectively.

\section{Discussions}
\label{sec:discuss}

Although the inferior results of the regression $p(\btheta \mid \bary)$ are best explained by insufficient data, we also identify a few possible reasons.
Our simple regressor is trained on parameter loss only and will likely be beneficial by adding audio loss, such as MSS, during training \cite{han_learning_2024,peladeau2024blind} to match the content better.
In addition, averaging the predictions does not guarantee lower MSS and MLDR than the individual predictions; thus, its evaluation in \cref{tab:results} could be sub-optimal.
Our proposed method could be improved using the conditional prior $p(\btheta \mid \tildex)$ instead of the unconditional one.
However, the dependency between the parameters and the raw vocal is apparently less than the one between the parameters and the processed audio.
Alternatively, we could even use $p(\btheta \mid \bary)$ as a stronger prior, which might substantially improve the performance.

Our work is limited by the assumption of fixed and ordered dimensionality in modelling the prior, which implies that the effects and routings are also fixed.
The actual prior must have variable dimensionality \cite{lee2023blind} and lots of equivariance in the parameters \cite{hayes_responsibility_2023, hayes_equivariant_2025}, e.g., the order of applying the effects does not matter, which is non-trivial to model.
In addition, our method requires differentiable effects, but the general idea should apply to non-differentiable ones \cite{christian_j_steinmetz_2024_14877423}, as long as the prior information is injected during optimisation.
\section{Conclusion}
\label{sec:conclusion}

In this paper, we proposed a calibrated ST-ITO by introducing a non-uniform Gaussian prior over the parameter space.
This calibration addresses the limitations of the original ST-ITO, which neglects prior information and yields suboptimal results.
We demonstrated the effectiveness of the proposed method on vocal effects style transfer, showing significant improvements in objective evaluations compared to baseline methods, including a blind effects estimator, nearest-neighbour approaches, and the original ST-ITO.
Our subjective evaluation shows a preference for the proposed method with the AFx-Rep encoder.
The results highlight the importance of incorporating prior knowledge into inference-time optimisation for better performance.

Future work will extend the proposed method to handle more complex effects configurations with variable dimensionality and equivariance in the parameter space.
Additionally, we aim to explore stronger priors, such as conditional priors dependent on raw vocals or reference tracks, and extend the method to non-differentiable effects.

\clearpage
\bibliographystyle{IEEEtran}
\bibliography{refs25}

\end{document}